\documentclass[twoside]{article}
\usepackage{Proc_NTSE}

\usepackage{graphicx}
\usepackage{amsmath}
\usepackage{amsfonts}
\usepackage{amssymb}
\newcommand{\gea}{\raisebox{-.3ex}{\small $ \
\stackrel{\textstyle >}{\sim} $ }}

\newcommand{\beq}{\begin{equation}}
\newcommand{\eeq}{\end{equation}}
\newcommand{\beqa}{\begin{eqnarray}}
\newcommand{\eeqa}{\end{eqnarray}}

\begin{document}
\thispagestyle{plain}

\begin{center}
{\Large \bf \strut
Origin and properties of strong inter-nucleon interactions\footnote{Dedicated to James Vary
on the occasion of his 70th birthday.}
\strut}\\
\vspace{10mm}
{\large \bf 
R. Machleidt}
\end{center}

\noindent{
\small\it Department of Physics, University of Idaho, Moscow, Idaho 83844, USA} \\

\markboth{R. Machleidt
}
{
Strong Inter-Nucleon Interactions} 

\begin{abstract}
I start with a historical review of the attempts to construct theories for the origin
of nuclear forces, for which I also summaries the most important properties.
The review then shifts to its main focus, which is the chiral effective
field theory approach to nuclear forces. I summarize the current status
of this approach and discuss the most important open issues:
the proper renormalization of the chiral
two-nucleon potential and sub-leading three-nucleon forces.
\\[\baselineskip] 
\end{abstract}

\section{Introduction and overview}

The nuclear force problem is as old as {\it James Vary}, namely seven decades.
What a coincidence!

The development of a proper theory of nuclear forces has occupied the minds of some of the brightest physicists and has been one of the main topics of physics research in the 20th century. The original idea was that the force is caused by the exchange of lighter particles (than nucleons) know as mesons, and this idea gave rise to the birth of a new sub-field of modern physics, namely, (elementary) particle physics. The modern perception of the nuclear force is that it is a residual interaction (similar to the van der Waals force between neutral atoms) of the even stronger force between quarks, which is mediated by the exchange of gluons and holds the quarks together inside a nucleon.

\subsection{Early history}
After the discovery of the neutron in 1932, it was clear that the atomic nucleus is made up from protons and neutrons. In such a system, electromagnetic forces cannot be the reason why the constituents of the nucleus are sticking together. Therefore, the concept of a new strong nuclear interaction was introduced. In 1935, the first theory for this new force was developed by the Japanese physicist Yukawa~\cite{Yuk35}, who suggested that the nucleons would exchange particles between each other and this mechanism would create the force. Yukawa constructed his theory in analogy to the theory of the electromagnetic interaction where the exchange of a (massless) photon is the cause of the force. However, in the case of the nuclear force, Yukawa assumed that the ``force-makers'' (which were eventually called ``mesons'') carry a mass of a fraction of the nucleon mass. This would limit the effect of the force to a finite range, since the uncertainty principal allows massive particles to travel only a finite distance. The meson predicted by Yukawa was finally found in 1947 in cosmic ray and in 1948 in the laboratory and called the pion. Yukawa was awarded the Nobel Prize in 1949. In the 1950's and 60's more mesons were found in accelerator experiments and the meson theory of nuclear forces was extended to include many mesons. These models became know as one-boson-exchange models, which is a reference to the fact that the different mesons are exchanged singly in this model. The one-boson-exchange model is very successful in explaining essentially all properties of the nucleon-nucleon interaction at low energies~\cite{BS64,HM75,Mac89,Mac01,DMS03}. In the 1970's and 80's, also meson models were developed that went beyond the simple single-particle exchange mechanism. These models included, in particular, the explicit exchange of two pions with all its complications. Well-known representatives of the latter kind are the Paris~\cite{Lac80} and the Bonn potential~\cite{MHE87}.

Since these meson models were quantitatively very successful, it appeared that they were the solution of the nuclear force problem. However, with the discovery (in the 1970's) that the fundamental theory of strong interactions is quantum chromo-dynamics (QCD) and not meson theory, all ``meson theories'' had to be viewed as models, and the attempts to derive a proper theory of the nuclear force had to start all over again.

\subsection{QCD and the nuclear force}
The problem with a derivation of nuclear forces from QCD is two-fold. First, each nucleon consists of three quarks such that the system of two nucleons is already a six-body problem. Second, the force between quarks, which is created by the exchange of gluons, has the feature of being very strong at the low energy-scale that is characteristic of nuclear physics. This extraordinary strength makes it difficult to find ``converging'' mathematical solutions. Therefore, during the first round of new attempts, QCD-inspired quark models became popular. The positive aspect of these models is that they try to explain nucleon structure (which consists of three quarks) and nucleon-nucleon interactions (six-quark systems) on an equal footing. Some of the gross features of the two-nucleon force, like the ``hard core'' are explained successfully in such models. However, from a critical point of view, it must be noted that these quark-based approaches are yet another set of models and not a theory. Alternatively, one may try to solve the six-quark problem with brute computing power, by putting the six-quark system on a four dimensional lattice of discrete points which represents the three dimensions of space and one dimension of time. This method has become known as lattice QCD and is making progress. However, such calculations are computationally very expensive and cannot be used as a standard nuclear physics tool.

\subsection{Chiral effective field theory}
Around 1990, a major breakthrough occurred when the nobel laureate Steven Weinberg applied the concept of an effective field theory (EFT) to low-energy QCD~\cite{Wei79,Wei91}. He simply wrote down the most general theory that is consistent with all the properties of low-energy QCD, since that would make this theory identical to low-energy QCD. A particularly important property is the so-called chiral symmetry, which is ``spontaneously'' broken. Massless particles observe chiral symmetry, which means that their spin and momentum are either parallel (``right-handed'') or anti-parallel (``left-handed'') and remain so forever. Since the quarks, which nucleons are made of (``up'' and ``down'' quarks), are almost mass-less, approximate chiral symmetry is a given. Naively, this symmetry should have the consequence that one finds in nature mesons of the same mass, but with positive and negative parity. However, this is not the case and such failure is termed a ``spontaneous'' breaking of the symmetry. According to a theorem first proven by Goldstone, the spontaneous breaking of a symmetry creates a particle, here, the pion. Thus, the pion becomes the main player in the production of the nuclear force. The interaction of pions with nucleons is weak as compared to the interaction of gluons with quarks. Therefore, pion-nucleon processes can be calculated without problem. Moreover, this effective field theory can be expanded in powers of momentum/scale, where ``scale'' denotes the ``chiral symmetry breaking scale'' which is about 1 GeV. This scheme is also known as chiral perturbation theory (ChPT) and allows to calculate the various terms that make up the nuclear force systematically power by power, or order by order. Another advantage of the chiral EFT approach is its ability to generate not only the force between two nucleons, but also many-nucleon forces, on the same footing~\cite{Wei92}.
In modern theoretical nuclear physics, the chiral EFT approach is becoming increasingly popular and is applied with great success~\cite{EHM09,ME11}.

\subsection{Main properties of the nuclear force and phenomenological potentials}
Some properties of nuclear interactions can be deduced from the properties of nuclei. The property of saturation suggests that nuclear forces are of short range (a few fm) and strongly attractive at that range, which explains nuclear binding. But the nuclear force has also a very complex spin-dependence. First evidence came from the observation that the deuteron (proton-neutron bound state, smallest atomic nucleus) deviates slightly from a spherical shape. This suggests a force that depends on the orientation of the spins of the nucleons with regard to the line connecting the two nucleons (tensor force). In heavier nuclei, a shell structure has been observed which according to a suggestion by Mayer and Jensen can be explained by a strong force between the spin of the nucleon and its orbital motion (spin-orbit force). More clear-cut evidence for the spin-dependence is extracted from scattering experiments where one nucleon is scattered off another nucleon. In such experiments, the existence of the nuclear spin-orbit and tensor forces has clearly been established. Scattering experiments at higher energies (more than 200 MeV) show indications that the nucleon-nucleon interaction at very short distances (smaller than 0.5 fm) becomes repulsive (``hard core''). 
Besides the force between two nucleons (2NF), there are also three-nucleon forces (3NF), four-nucleon forces (4NF), etc. However, the 2NF is much stronger than the 3NF, which in turn is much stronger than the 4NF, etc. In exact calculations of the properties of light nuclei based upon the bare nuclear forces, it has been shown that 3NFs are important. Their contribution is small, but crucial. The need for 4NF for explaining nuclear properties has not (yet) been clearly established.

Phenomenological nucleon-nucleon ($NN$) potentials are constructed is close
relationship to the empirical facts. In this regard, the most faithful method of construction
is inverse scattering theory, which the so-called JISP-16 potentials are based upon~\cite{Shi09}.
\\
\\
In the following sections, I will elaborate more on the theory of nuclear forces with particular emphasis on the view according to which the forces between nucleons emerge from low-energy QCD
via an effective field theory.

\section{Effective field theory for low-energy QCD
\label{sec_EFT}}

Quantum chromodynamics (QCD) is the theory of strong interactions.
It deals with quarks, gluons and their interactions and is
part of the Standard Model of Particle Physics.
QCD is a non-Abelian gauge field theory
with color $SU(3)$ the underlying gauge group.
The non-Abelian nature of the theory has dramatic
consequences. While 
the interaction between colored objects is weak 
at short distances or high momentum transfer
(``asymptotic freedom'');
it is strong at long distances ($\gea 1$ fm) or low energies,
leading to the confinement of quarks into colorless
objects, the hadrons. Consequently, QCD allows for a 
perturbative analysis at large energies, whereas it is
highly non-perturbative in the low-energy regime.
Nuclear physics resides at low energies and
the force between nucleons is
a residual color interaction
similar to the van der Waals force between neutral molecules.
Therefore, in terms of quarks and gluons, the nuclear force
is a very complicated problem that, nevertheless, can be attacked
with brute computing power on a discretized, Euclidean space-time lattice
(known as lattice QCD). In a recent study~\cite{Bea06}, the neutron-proton scattering lengths
in the singlet and triplet $S$-waves have been determined in fully dynamical
lattice QCD. This result is then extrapolated
to the physical pion mass with the help of chiral perturbation theory. The pion mass
of 354 MeV is still too large to allow for reliable extrapolations, but the feasibility has been
demonstrated and more progress can be expected for the near future.
In a lattice calculation of a very different kind, the $NN$ potential
was studied~\cite{Hat12}. The central part of the
potential shows a repulsive core plus attraction of intermediate range. This is a very promising
result, but it must be noted that also in this investigation still rather large pion masses are being used.
In any case, advanced lattice QCD calculations are under way and continuously improved. 
However, since these calculations are very time-consuming
and expensive, they can only be used to check a few representative key-issues. For everyday
nuclear structure physics, a more efficient approach is needed. 

The efficient approach is an effective field theory.
For the development of an EFT, it is crucial to identify a separation of
scales. In the hadron spectrum, a large gap between the masses of
the pions and the masses of the vector mesons, like $\rho(770)$ and $\omega(782)$,
can clearly be identified. Thus, it is natural to assume that the pion mass sets the soft scale, 
$Q \sim m_\pi$,
and the rho mass the hard scale, $\Lambda_\chi \sim m_\rho$, also known
as the chiral-symmetry breaking scale.
This is suggestive of considering an expansion in terms of the soft scale over the hard scale,
$Q/\Lambda_\chi$.
Concerning the relevant degrees of freedom, we noticed already that,
for the ground state and the
low-energy excitation spectrum of
an atomic nucleus as well as for conventional nuclear
reactions,
quarks and gluons are ineffective degrees of freedom,
while nucleons and pions are the appropriate ones.
To make sure that this EFT is not just another phenomenology,
it must have a firm link with QCD.
The link is established by having the EFT observe
all relevant symmetries of the underlying theory.
This requirement is based upon a `folk theorem' by
Weinberg~\cite{Wei79}:
\begin{quote}
If one writes down the most general possible Lagrangian, including {\it all}
terms consistent with assumed symmetry principles,
and then calculates matrix elements with this Lagrangian to any given order of
perturbation theory, the result will simply be the most general possible 
S-matrix consistent with analyticity, perturbative unitarity,
cluster decomposition, and the assumed symmetry principles.
\end{quote}
In summary, the EFT program consists of the following steps:
\begin{enumerate}
\item
Identify the soft and hard scales, and the degrees of freedom (DOF) appropriate
for (low-energy) nuclear physics. Soft scale: $Q \sim m_\pi$, hard scale:
$\Lambda_\chi \sim m_\rho \sim 1$ GeV; DOF: pions and nucleons.
\item
Identify the relevant symmetries of low-energy QCD and
investigate if and how they are broken: explicitly and spontaneously
broken chiral symmetry (spontaneous symmetry breaking generates the pions
as Goldstone bosons).
\item
Construct the most general Lagrangian consistent with those
symmetries and symmetry breakings, see Ref.~\cite{ME11}.
\item
Design an organizational scheme that can distinguish
between more and less important contributions: 
a low-momentum expansion, $(Q/\Lambda_\chi )^\nu$, with $\nu$ determined
by `power counting'. For an irreducible diagram that involves $A$ nucleons, we have:
\begin{equation} \nu = -2 +2A - 2C + 2L 
+ \sum_i \Delta_i \, .  
\label{eq_nu} 
\end{equation}
where
\begin{equation}
\Delta_i  \equiv   d_i + \frac{n_i}{2} - 2  \, ,
\label{eq_Deltai}
\end{equation}
with $C$ the number of separately connected pieces and
$L$ the number of loops in the diagram;
$d_i$ is the number of derivatives or pion-mass insertions 
and $n_i$ the number of nucleon fields (nucleon legs)
involved in vertex $i$;
the sum runs over all vertices $i$ contained in the diagram 
under consideration.
Note that for an irreducible 
$NN$ diagram ($A=2$, $C=1$), the
power formula collapses to the very simple expression
\begin{equation}
\nu =  2L + \sum_i \Delta_i \,.
\label{eq_nunn}
\end{equation}
\item
Guided by the expansion, calculate Feynman diagrams
for the problem under consideration
to the desired accuracy (see next Section).
\end{enumerate}

\section{The hierarchy of nuclear forces in chiral EFT}

\begin{figure}[t]\centering
\vspace*{-0.9cm}
\scalebox{0.55}{\includegraphics{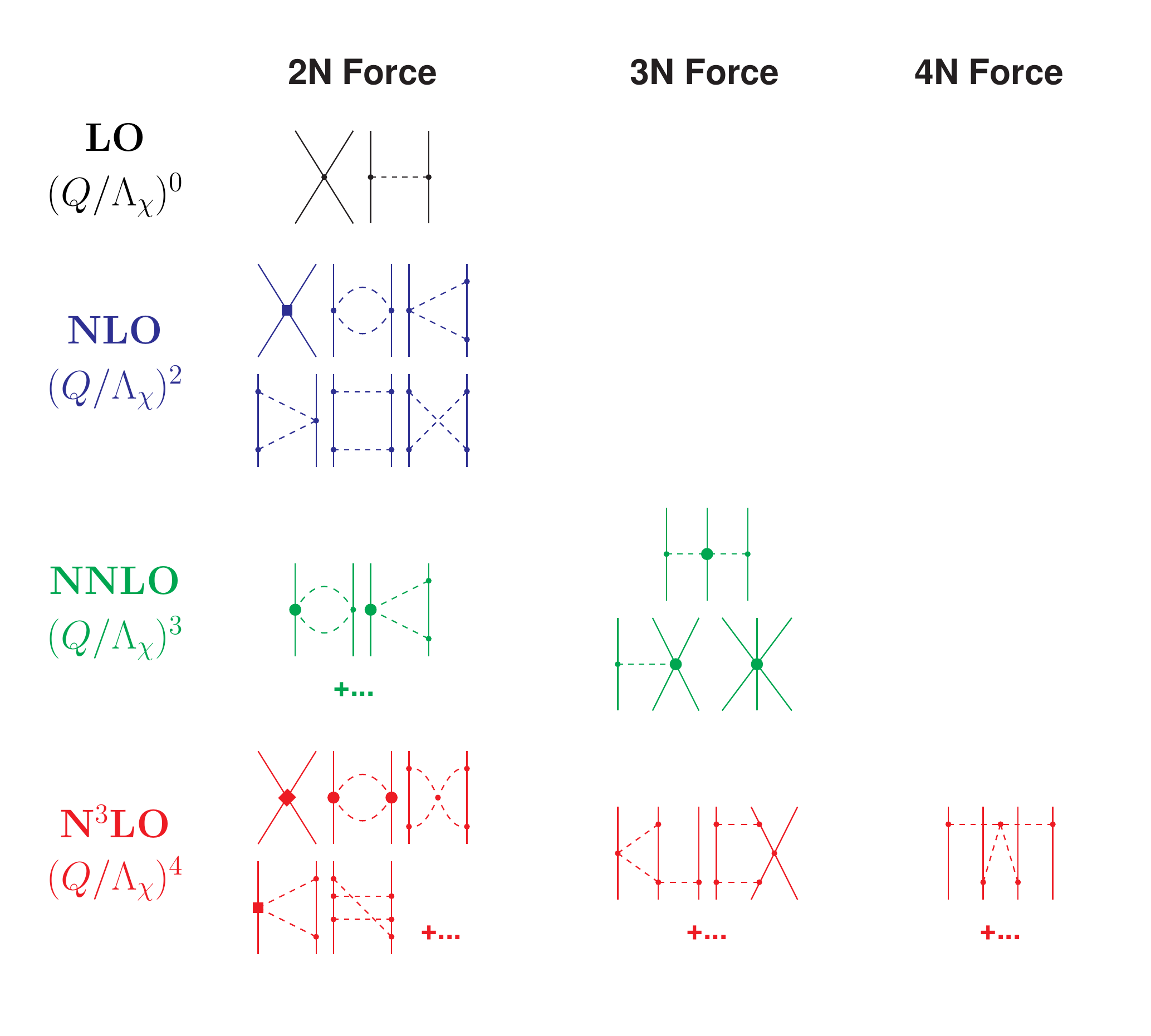}}
\vspace*{-1.0cm}
\caption{Hierarchy of nuclear forces in ChPT. Solid lines
represent nucleons and dashed lines pions. 
Small dots, large solid dots, solid squares, and solid diamonds
denote vertices of index $\Delta_i= \, $ 0, 1, 2, and 4, respectively. 
Further explanations are
given in the text.}
\label{fig_hi}
\end{figure}

Chiral perturbation theory and power counting
imply that nuclear forces emerge as a hierarchy
controlled by the power $\nu$, Fig.~\ref{fig_hi}.

In lowest order, better known as leading order (LO, $\nu = 0$), 
the $NN$ amplitude
is made up by two momentum-independent contact terms
($\sim Q^0$), 
represented by the 
four-nucleon-leg graph
with a small-dot vertex shown in the first row of 
Fig.~\ref{fig_hi},
and
static one-pion exchange (1PE), second
diagram in the first row of the figure.
This is, of course, a rather rough approximation
to the two-nucleon force (2NF), but accounts already for some
important features.
The 1PE provides the tensor force,
necessary to describe the deuteron, and it explains
$NN$ scattering in peripheral partial waves of very high
orbital angular momentum. At this order, the two contacts 
which contribute only in $S$-waves provide
the short- and intermediate-range interaction which is somewhat
crude.

In the next order,
$\nu=1$, all contributions vanish due to parity
and time-reversal invariance.

Therefore, the next-to-leading order (NLO) is $\nu=2$.
Two-pion exchange (2PE) occurs for the first time
(``leading 2PE'') and, thus, the creation of
a more sophisticated
description of the intermediate-range interaction
is starting here. 
Since the loop involved in each pion-diagram implies
already $\nu=2$ [cf.\ Eq.~(\ref{eq_nunn})],
the vertices must have $\Delta_i = 0$.
Therefore, at this order, only the lowest order
$\pi NN$ and $\pi \pi NN$ vertices are allowed which
is why the leading 2PE is rather weak.
Furthermore, there are 
seven contact terms of 
${\cal O}(Q^2)$, 
shown by
the four-nucleon-leg graph with a solid square,
 which contribute
in $S$ and $P$ waves. The operator structure of these
contacts include a spin-orbit term besides central,
spin-spin, and tensor terms. Thus, essentially all spin-isospin
structures necessary to describe the two-nucleon
force phenomenologically have been generated at this order.
The main deficiency at this stage of development 
is an insufficient intermediate-range attraction.

This problem is finally fixed at order three 
($\nu=3$), next-to-next-to-leading order (NNLO).
The 2PE involves now the two-derivative
$\pi\pi NN$ seagull vertices (proportional to
the $c_i$ LECs) denoted by a large solid dot
in Fig.~\ref{fig_hi}.
These vertices represent correlated 2PE
as well as intermediate $\Delta(1232)$-isobar contributions.
It is well-known from the meson phenomenology of 
nuclear forces~\cite{Lac80,MHE87}
that these two contributions are crucial
for a realistic and quantitative 2PE model.
Consequently, the 2PE now assumes a realistic size
and describes the intermediate-range attraction of the
nuclear force about right. Moreover, first relativistic 
corrections come into play at this order.
There are no new contacts.

The reason why we talk of a hierarchy of nuclear forces is that 
two- and many-nucleon forces are created on an equal footing
and emerge in increasing number as we go to higher and higher orders.
At NNLO, the first set of
nonvanishing three-nucleon forces (3NF) occur~\cite{Kol94,Epe02b},
cf.\ column `3N Force' of
Fig.~\ref{fig_hi}. 
In fact, at the previous order, NLO,
irreducible 3N graphs appear already, however,
it has been shown by Weinberg~\cite{Wei92}  that these diagrams all cancel.
Since nonvanishing 3NF contributions happen first
at order 
$(Q/\Lambda_\chi)^3$, 
they are very weak as compared to 2NF which start at
$(Q/\Lambda_\chi)^0$.

\begin{figure}[t]
\vspace*{-1.5cm}
\hspace*{-1.6cm}
\scalebox{0.41}{\includegraphics{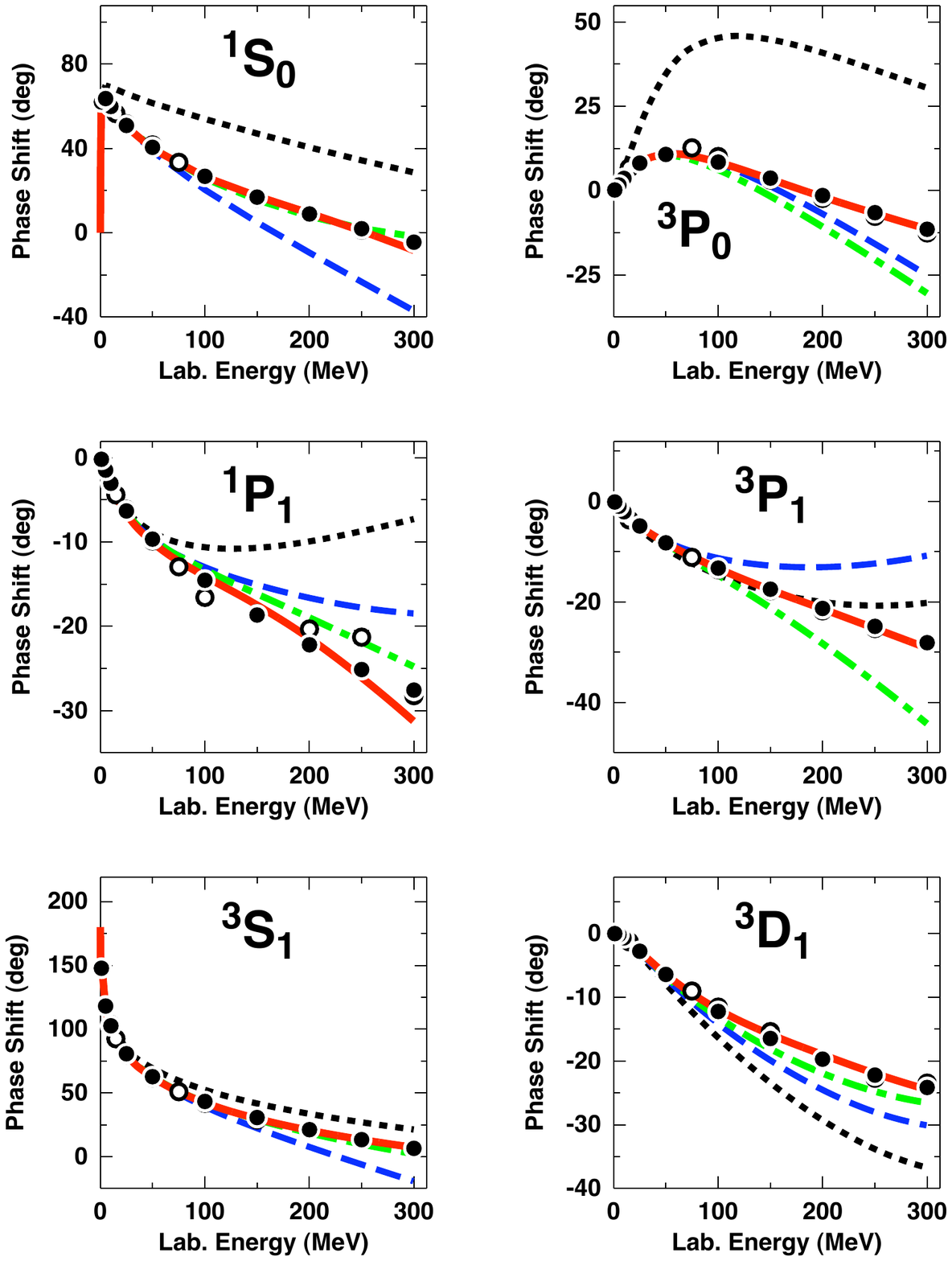}}
\hspace*{-2.3cm}
\scalebox{0.41}{\includegraphics{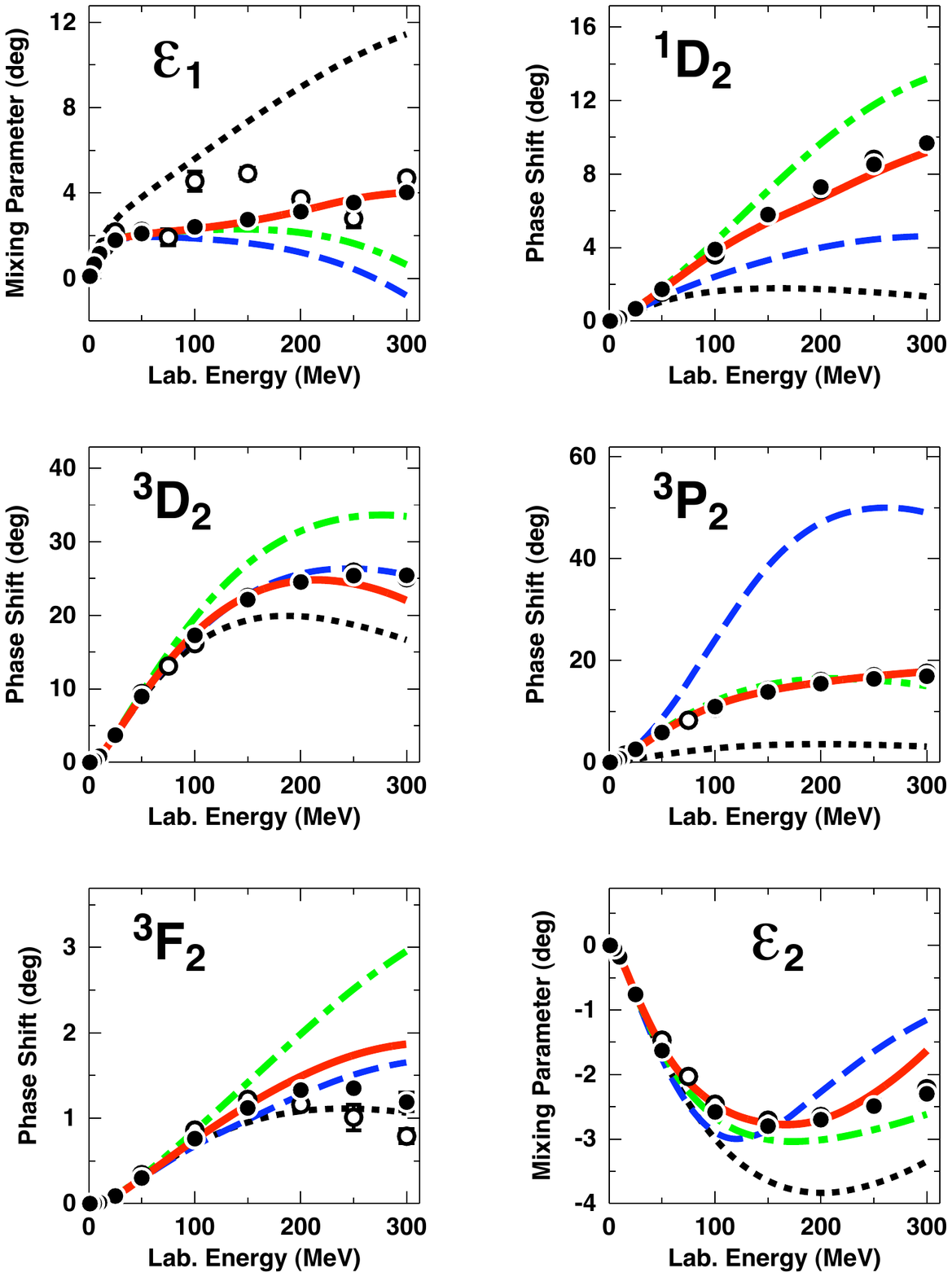}}
\vspace*{-2.3cm}
\caption{Phase shifts of $np$ scattering
as calculated from $NN$ potentials at different
orders of ChPT. The black dotted line is LO(500), 
the blue dashed is NLO(550/700)~\cite{EGM04},
the green dash-dotted NNLO(600/700)~\cite{EGM04},
and the red solid N$^3$LO(500)~\cite{EM03}, where the 
numbers in parentheses denote the cutoffs in MeV.
Partial waves with total angular momentum $J\leq 2$
are displayed.
The solid dots and open circles are the results from the Nijmegen
multi-energy $np$ phase shift analysis~\protect\cite{Sto93} 
and the VPI/GWU
single-energy $np$ analysis SM99~\protect\cite{SM99}, respectively.
\label{fig_phorders}}
\end{figure}

More 2PE is produced at $\nu =4$, next-to-next-to-next-to-leading
order (N$^3$LO), of which we show only a few symbolic diagrams in 
Fig.~\ref{fig_hi}. 
Two-loop 2PE
graphs show up for the first time and so does
three-pion exchange (3PE) which necessarily involves
two loops.
3PE was found to be negligible at this order~\cite{Kai00a,Kai00b}.
Most importantly, 15 new contact terms $\sim Q^4$
arise and are represented 
by the four-nucleon-leg graph with a solid diamond.
They include a quadratic spin-orbit term and
contribute up to $D$-waves.
Mainly due to the increased number of contact terms,
a quantitative description of the
two-nucleon interaction up to about 300 MeV
lab.\ energy is possible, 
at N$^3$LO 
(see red solid line in Fig.~\ref{fig_phorders} and cf.\ Table~\ref{tab_chi2}).
Besides further 3NF,
four-nucleon forces (4NF) start
at this order. Since the leading 4NF 
come into existence one
order higher than the leading 3NF, 4NF are weaker
than 3NF.
Thus, ChPT provides a straightforward explanation for
the empirically known fact that 2NF $\gg$ 3NF $\gg$ 4NF
\ldots.

\begin{table}[b]
\caption{
$\chi^2$/datum for the reproduction of the 1999 $np$ 
database~\protect\cite{note2} below 290 MeV by various $np$ potentials.
$T_{\rm lab}$ denotes the kinetic energy of the incident neutron
in the laboratory system.
\label{tab_chi2}}
\smallskip
\begin{tabular*}{\textwidth}{@{\extracolsep{\fill}}cccccc}
\hline 
\noalign{\smallskip}
 $T_{\rm lab}$ bin (MeV)
 & \# of $np$ data 
 & N$^3$LO~\cite{EM03}
 & NNLO~\cite{EGM04}
 & NLO~\cite{EGM04}
 & AV18~\cite{WSS95}
\\
\hline 
\noalign{\smallskip}
0--100&1058&1.05&1.7&4.5&0.95\\
100--190&501&1.08&22&100&1.10\\
190--290&843&1.15&47&180&1.11\\
\hline
\noalign{\smallskip}
0--290&2402&1.10&20&86&1.04
\\ 
\hline 
\end{tabular*}
\end{table}

During the past decade or so,
chiral two-nucleon forces have been used in many microscopic calculations of
nuclear reactions and structure~\cite{DF07,Cor02,Cor05,Cor10,NC04,Hag08,FOS09} 
and the combination of chiral two- and three-nucleon forces has been applied in
few-nucleon reactions~\cite{Epe02b,Wit06,Kal12,Kie10,Viv10},
structure of light- and medium-mass 
nuclei~\cite{Nog06,Nav07,Ots09,HKW09,Rot12,Her13,Hag12a,Hag12b,Eks13},
and nuclear and neutron matter~\cite{HS10,Heb11,Sam12,Cor12}---with a great deal of success.
The majority of nuclear structure calculations is nowadays based upon chiral forces.

However, in spite of this progress, we are not done. Due to the complexity of the
nuclear force issue, there are still many subtle and not so subtle open problems.
We will not list and discuss all of them, but instead just focus on the two open issues, 
which we perceive as the most important ones:
\begin{itemize}
\item
The proper renormalization of chiral nuclear potentials and
\item
Subleading chiral few-nucleon forces.
\end{itemize}

\section{Renormalization of chiral nuclear forces \label{sec_reno}}

\subsection{The chiral $NN$ potential}
In mathematical terms,
the various orders of the irreducible graphs in Fig.~\ref{fig_hi},
 which define the chiral $NN$ potential, 
are given by:
\beqa
V_{\rm LO} & = & 
V_{\rm ct}^{(0)} + 
V_{1\pi}^{(0)} 
\label{eq_VLO}
\\
V_{\rm NLO} & = & V_{\rm LO} +
V_{\rm ct}^{(2)} + 
V_{1\pi}^{(2)} +
V_{2\pi}^{(2)} 
\label{eq_VNLO}
\\
V_{\rm NNLO} & = & V_{\rm NLO} +
V_{1\pi}^{(3)} + 
V_{2\pi}^{(3)} 
\label{eq_VNNLO}
\\
V_{{\rm N}^3{\rm LO}} & = & V_{\rm NNLO} +
V_{\rm ct}^{(4)} +
V_{1\pi}^{(4)} +  
V_{2\pi}^{(4)} +
V_{3\pi}^{(4)} 
\label{eq_VN3LO}
\eeqa
where 
the superscript denotes the order $\nu$ of the low-momentum
expansion.
Contact potentials carry the subscript ``ct'' and
pion-exchange potentials can be identified by an
obvious subscript.

Multi-pion exchange, which starts at NLO and continues through
all higher orders, involves
divergent loop integrals that need to be regularized.
An elegant way to do this is dimensional regularization
which 
(besides the main nonpolynomial result) 
typically generates polynomial terms with coefficients
that are, in part, infinite or scale dependent.
One purpose of the contacts is
to absorb all infinities and scale dependencies and make
sure that the final result is finite and scale independent.
This is the renormalization of the perturbatively calculated
$NN$ amplitude (which, by definition, is the ``$NN$ potential'').
It is very similar to what is done in the ChPT calculations
of $\pi\pi$ and $\pi N$ scattering, namely, a renormalization
order by order, which is the method of choice for any EFT.
Thus, up to this point, the calculation fully meets the
standards of an EFT and there are no problems.
The perturbative $NN$ amplitude can be used to make model
independent predictions for peripheral partial waves.

\subsection{Nonperturbative renormalization of the $NN$ potential}
For calculations of the structure of nuclear few and many-body systems,
the lower partial waves are the most important ones. The fact that
in $S$ waves we have large scattering lengths and shallow (quasi)
bound states indicates that these waves need to be treated nonperturbatively.
Following Weinberg's prescription~\cite{Wei91}, this is accomplished by
inserting the potential $V$ into the Lippmann-Schwinger (LS) equation:
\begin{equation}
 {T}({\vec p}~',{\vec p})= {V}({\vec p}~',{\vec p})+
\int d^3p''\:
{V}({\vec p}~',{\vec p}~'')\:
\frac{M_N}
{{ p}^{2}-{p''}^{2}+i\epsilon}\:
{T}({\vec p}~'',{\vec p}) \,,
\label{eq_LS}
\end{equation}
where $M_N$ denotes the nucleon mass.

In general, the integral in
the LS equation is divergent and needs to be regularized.
One way to do this is  by
multiplying $V$
with a regulator function
\begin{equation}
{ V}(\vec{ p}~',{\vec p}) 
\longmapsto
{ V}(\vec{ p}~',{\vec p})
\;\mbox{\boldmath $e$}^{-(p'/\Lambda)^{2n}}
\;\mbox{\boldmath $e$}^{-(p/\Lambda)^{2n}}
\label{eq_regulator} \,.
\end{equation}
Typical choices for the cutoff parameter $\Lambda$ that
appears in the regulator are 
$\Lambda \approx 0.5 \mbox{ GeV} < \Lambda_\chi \approx 1$ GeV.

It is pretty obvious that results for the $T$-matrix may
depend sensitively on the regulator and its cutoff parameter.
This is acceptable if one wishes to build models.
For example, the meson models of the past~\cite{Mac89}
always depended sensitively on the choices for the
cutoff parameters which, in fact,
were important for the fit of the $NN$ data.
However, the EFT approach wishes to be fundamental
in nature and not just another model.

In field theories, divergent integrals are not uncommon and methods have
been developed for how to deal with them.
One regulates the integrals and then removes the dependence
on the regularization parameters (scales, cutoffs)
by renormalization. In the end, the theory and its
predictions do not depend on cutoffs
or renormalization scales.
So-called renormalizable quantum field theories, like QED,
have essentially one set of prescriptions 
that takes care of renormalization through all orders. 
In contrast, 
EFTs are renormalized order by order. 

Weinberg's implicit assumption~\cite{Wei91,Wei09} was that the counterterms
introduced to renormalize the perturbatively calculated
potential, based upon naive dimensional analysis (``Weinberg counting''),
are also sufficient to renormalize the nonperturbative
resummation of the potential in the LS equation.
In 1996, Kaplan, Savage, and Wise (KSW)~\cite{KSW96}
pointed out that there are problems with the Weinberg scheme
if the LS equation is renormalized 
by minimally-subtracted dimensional regularization.
This criticism resulted in a flurry of publications on
the renormalization of the nonperturbative
$NN$ problem.
The literature is too comprehensive
to elaborate on all contributions. Therefore, we will restrict ourselves, here, to discussing
just a few aspects that we perceive as particularly important. 
A more comprehensive consideration can be found in Ref.~\cite{ME11}

Naively, the most perfect renormalization procedure is the one where the cutoff
parameter $\Lambda$ is carried to infinity while stable results are maintained.
This was done successfully at LO in the work by Nogga {\it et al}~\cite{NTK05}.
At NNLO, the infinite-cutoff renormalization procedure has been investigated 
in~\cite{YEP07} for partial waves with total angular momentum $J\leq 1$ and
in~\cite{VA07} for all partial waves with $J\leq 5$. At N$^3$LO, the $^1S_0$ 
state was considered in Ref.~\cite{Ent08}, and all states up to $J=6$ were investigated in 
Ref.~\cite{ZME12}.
From all of these works, it is evident that no counter term is effective in partial-waves with
short-range repulsion and only a single counter term can effectively be used in
partial-waves with short-range attraction. Thus, for the $\Lambda \rightarrow \infty$
renormalization prescription, even at N$^3$LO, there exists either one or no counter term
per partial-wave state. This is inconsistent with any reasonable power-counting scheme
and prevents an order-by-order improvement of the predictions.

To summarize:
In the infinite-cutoff renormalization scheme, the potential is admitted up to unlimited momenta.  However, the EFT this potential is derived from has validity only for momenta smaller than the chiral symmetry breaking scale $\Lambda_{\chi}\approx 1$ GeV.  The lack of order-by-order convergence and discrepancies in lower partial-waves demonstrate that the potential should not be used beyond the limits of the effective theory~\cite{ZME12} (see Ref.~\cite{EG09}
for a related discussion).  The conclusion then is that cutoffs should be limited to $\Lambda\lesssim\Lambda_{\chi}$
(but see also Ref.~\cite{EG12}).

A possible solution of this problem was proposed already in~\cite{NTK05}
and reiterated in a paper by Long and van Kolck~\cite{LK08}.
A calculation of the proposed kind has been performed by
Valderrama~\cite{Val11}, for the $S$, $P$, and $D$ waves. 
The author renormalizes the LO interaction nonperturbatively
and then uses the LO
distorted wave to calculate the 2PE contributions at NLO and NNLO
perturbatively. It turns out that perturbative renormalizability requires the introduction
of about twice as many counter terms as compared to Weinberg counting, 
which reduces the predictive power. The order-by-order convergence of 
the $NN$ phase shifts appears to be reasonable.

\begin{figure*}
\vspace*{-3.0cm}
\includegraphics[scale=.5]{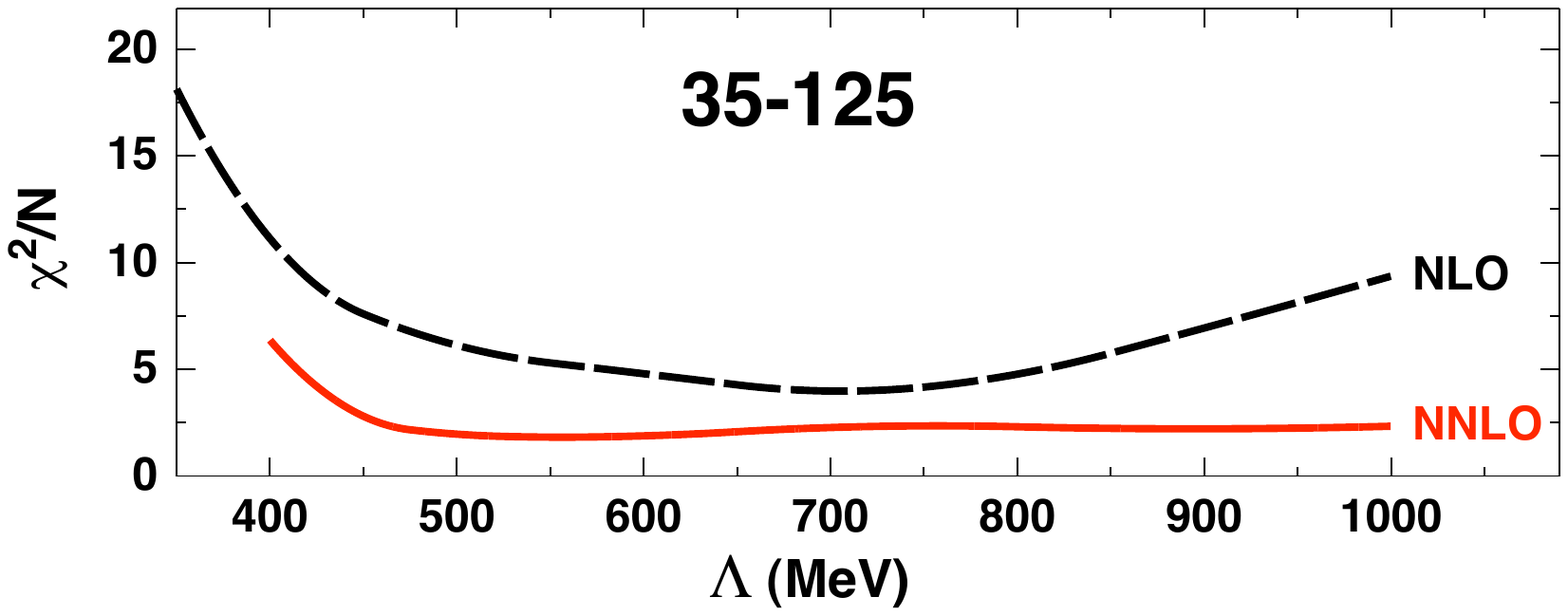}

\vspace*{-9.7cm}
\includegraphics[scale=.5]{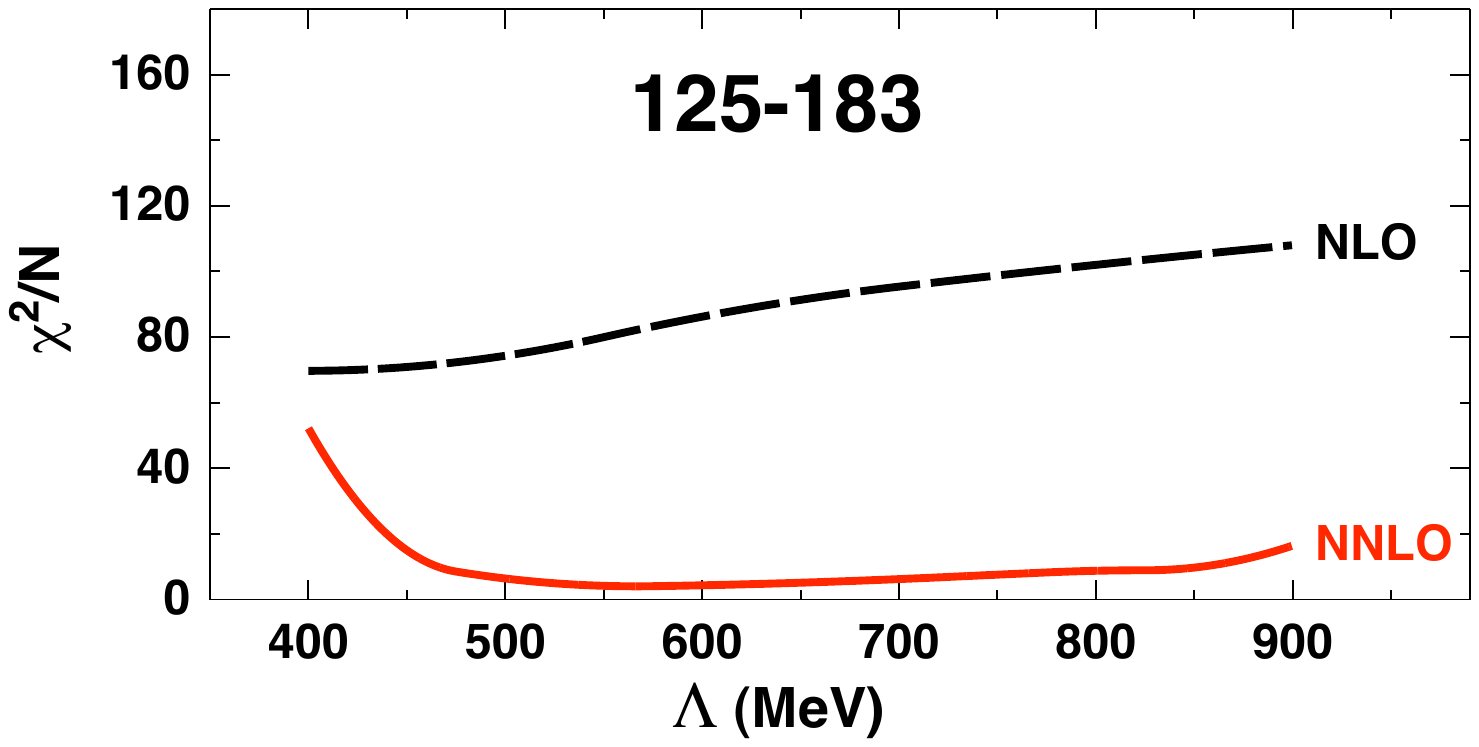}
\vspace*{-8.0cm}
\caption{$\chi^2$/datum for the reproduction of the $np$ data in the
energy range 35-125 MeV (upper frame) and 125-183 MeV (lower frame)
as a function of the cutoff parameter $\Lambda$ of the regulator function
Eq.~(\ref{eq_regulator}). The (black) dashed curves show the $\chi^2$/datum
achieved with $np$ potentials constructed at order NLO and the (red) solid curves are for NNLO.}
\label{fig_32}       
\end{figure*}

However, even if one considers the above method as successful for $NN$ scattering,
there is doubt if the interaction generated in this approach is of any use
for applications in nuclear few- and many-body problems.
In applications, one would first have to solve the many-body problem
with the re-summed LO interaction, and then add higher order corrections in
perturbation theory.
It was shown in a recent paper~\cite{Mac09} that the renormalized LO
interaction
is characterized by a very large tensor force from 1PE. This is no surprise since
LO is renormalized with $\Lambda \rightarrow \infty$ implying that the 1PE,
particulary its tensor force, is totally uncut.
As a consequence of this, the wound integral in nuclear matter, $\kappa$,
comes out to be about 40\%. The hole-line and coupled cluster expansions
are known to converge $\propto \kappa^{n-1}$
with $n$ the number of hole-lines or particles per cluster.
For conventional nuclear forces, the wound integral is typically between 5 and 10\%
and the inclusion of three-body clusters (or three hole-lines) are needed to
obtain converged results in the many-body system.
Thus, if the wound integral is 40\%, probably, up to six hole-lines need to be
included for approximate convergence. Such calculations are not feasible even with
the most powerful computers of today and will not be feasible any time soon.
Therefore, even if the renormalization procedure proposed in~\cite{LK08} will work
for $NN$ scattering, the interaction produced will be highly impractical (to say
the least) in applications in few- and many-body problems because of convergence problems
with the many-body energy and wave functions.

Crucial for an EFT are regulator independence (within the range of validity
of the EFT) and a power counting scheme that allows for order-by-order
improvement with decreasing truncation error.
The purpose of renormalization is to achieve this regulator independence while maintaining
a functional power counting scheme.

Thus, in the spirit of Lepage~\cite{Lep97}, the cutoff independence should be examined
for cutoffs below the hard scale and not beyond. Ranges of cutoff independence within the
theoretical error are to be identified using Lepage plots~\cite{Lep97}.
Recently, we have started a systematic investigation of this kind.
In our work, we quantify the error of the predictions by calculating the $\chi^2$/datum 
for the reproduction of the neutron-proton ($np$) elastic scattering data
as a function of the cutoff parameter $\Lambda$ of the regulator function
Eq.~(\ref{eq_regulator}). We have investigated the predictions by chiral $np$ potentials at 
order NLO and NNLO applying Weinberg counting for the counter terms ($NN$ contact terms).
We show our results for the energy range 35-125 MeV in the upper frame of Fig.~\ref{fig_32}
and for 125-183 MeV in the lower frame. 
It is seen that the reproduction of the $np$ data at these energies is generally poor
at NLO, while at NNLO the $\chi^2$/datum assumes acceptable values (a clear demonstration of
order-by-order improvement). Moreover, at NNLO one observes 
``plateaus'' of constant low $\chi^2$ for
cutoff parameters ranging from about 450 to 850 MeV. This may be perceived as cutoff independence
(and, thus, successful renormalization) for the relevant range of cutoff parameters.

\section{Few-nucleon forces and what is missing \label{sec_manyNF}}

We will now discuss the other issue we perceive as unfinished and important, namely,
subleading chiral few-nucleon forces.

Nuclear three-body forces in ChPT were initially discussed
by Weinberg~\cite{Wei92}.
The 3NF at NNLO, was derived by van Kolck~\cite{Kol94}
and applied, for the first time, in nucleon-deuteron
scattering by Epelbaum {\it et al.}~\cite{Epe02b}.
The leading 4NF (at N$^3$LO) was constructed by
Epelbaum~\cite{Epe06} and found to contribute in the
order of 0.1 MeV to the $^4$He binding energy
(total $^4$He binding energy: 28.3 MeV)
in a preliminary calculation~\cite{Roz06}, confirming the traditional
assumption that 4NF are essentially negligible.
Therefore, the focus is on 3NFs.

For a 3NF, we have $A=3$ and $C=1$ and, thus, Eq.~(\ref{eq_nu}) implies
\begin{equation}
\nu = 2 + 2L + 
\sum_i \Delta_i \,.
\label{eq_nu3nf}
\end{equation}
We will use this equation to analyze 3NF contributions
order by order.
The first non-vanishing 3NF occurs at $\nu=3$ (NNLO), which
is obtained when
there are no loops ($L=0$) and 
$\sum_i \Delta_i = 1$, i.e., 
$\Delta_i=1$ for one vertex 
while $\Delta_i=0$ for all other vertices.
There are three topologies which fulfill this condition,
known as the two-pion exchange (2PE), one-pion exchange (1PE),
and contact graphs (cf.\ Fig.~\ref{fig_hi}).

The 3NF at NNLO
has been applied in
calculations of few-nucleon reactions~\cite{Kal12},
structure of light- and medium-mass 
nuclei~\cite{Nog06,Nav07,Ots09,HKW09,Rot12,Her13,Hag12a,Hag12b,Eks13},
and nuclear and neutron matter~\cite{HS10,Heb11,Sam12,Cor12}
with a great deal of success.
However, the famous `$A_y$ puzzle' of nucleon-deuteron scattering~\cite{Epe02b} and 
the analogous problem with the
analyzing power in $p$-$^3$He scattering~\cite{Viv10} 
is not resolved.
Furthermore, the spectra of light nuclei leave room for improvement~\cite{Nav07}.
Since we are dealing with a perturbation theory, it is natural to turn to the next
order when looking for improvements.

The next order is N$^3$LO, where we have loop and tree diagrams.
For the loops, we have
$L=1$ and, therefore, all $\Delta_i$ have to be zero
to ensure $\nu=4$. 
Thus, these one-loop 3NF diagrams can include
only leading order vertices, the parameters of which
are fixed from $\pi N$ and $NN$ analysis.
One sub-group of these diagrams (the 2PE graphs)
has been calculated by Ishikawa and Robilotta~\cite{IR07},
and the other topologies
have been evaluated by the Bochum-Bonn group~\cite{Ber08,Ber11}.
The N$^3$LO 2PE 3NF has been applied in the calculation
of nucleon-deuteron observables in Ref.~\cite{IR07} 
causing little impact.
Very recently, the long-range part of the chiral N$^3$LO 3NF has been
tested in the triton~\cite{Ski11} and in three-nucleon scattering~\cite{Wit12}
yielding only moderate effects. The long- and short-range parts of this
force have been used in neutron matter calculations
(together with the N$^3$LO 4NF) producing relatively large contributions
from the 3NF~\cite{Tew12}. Thus, the ultimate assessment of the N$^3$LO 3NF is still
outstanding and will require more few- and many-body applications.

\begin{figure}[t]
\vspace*{-0.8cm}
\scalebox{0.9}{\includegraphics{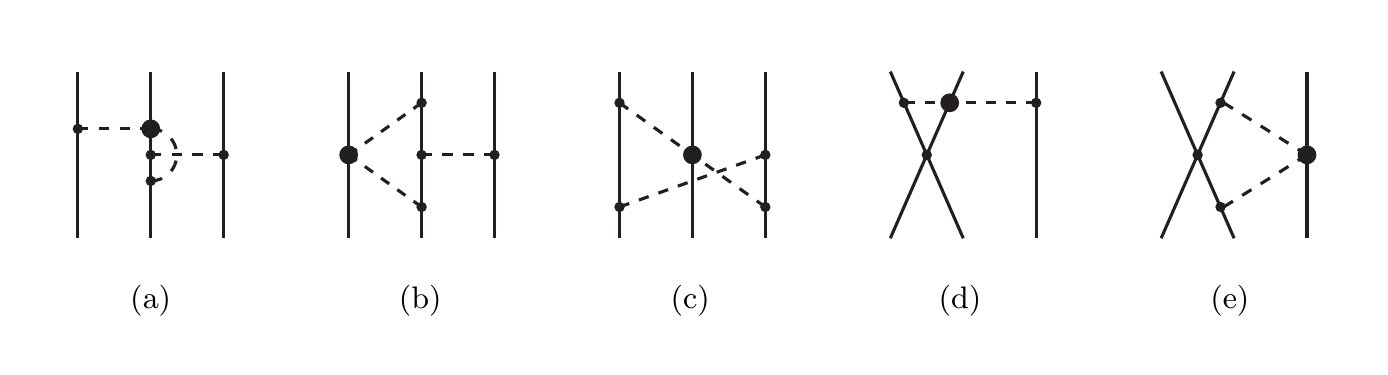}}
\vspace*{-0.8cm}
\caption{3NF one-loop contributions at N$^4$LO ($\nu=5$). We show one representative diagram for each of five
topologies, which are: (a) 2PE, (b) 2PE-1PE, (c) ring, (d) 1PE-contact, and (e) 2PE-contact. 
Notation as in Fig.~\ref{fig_hi}.}
\label{fig_3nfn4loloops}
\end{figure}

\begin{figure}[b]\center
\vspace*{-0.5cm}
\scalebox{0.9}{\includegraphics{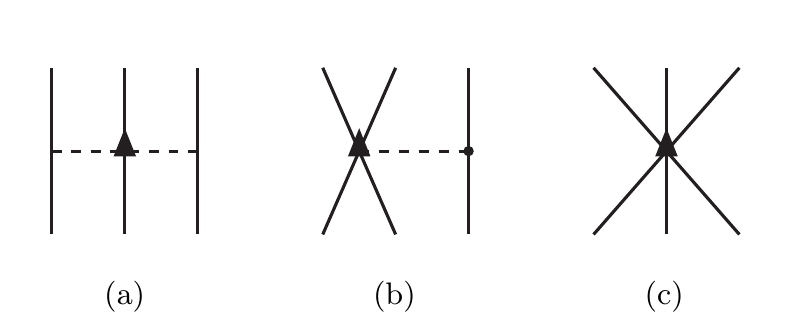}}
\vspace*{-0.5cm}
\caption{3NF tree graphs at N$^4$LO ($\nu=5$) denoted by: (a) 2PE, (b) 1PE-contact, and (c) contact. 
Solid triangles represent vertices of index $\Delta_i = 3$.}
\label{fig_3nfn4lotrees}
\end{figure}

In the meantime, it is of interest to take already a look
at the next order of 3NFs, which is N$^4$LO or $\nu=5$
(of the $\Delta$-less theory to which  the present discussion
is restricted because of lack of space). 
The loop contributions that occur at this order
are obtained by replacing in the N$^3$LO loops
one vertex by a $\Delta_i=1$ vertex (with LEC $c_i$), Fig.~\ref{fig_3nfn4loloops},
which is why these loops may be more sizable than the N$^3$LO loops.
The 2PE topology turns out to be of modest size~\cite{KGE12};
moreover, it can be handled in a practical way by summing it up
together with the 2PE topologies at NNLO and N$^3$LO~\cite{KGE12}.
The 2PE-1PE and ring topologies have also been derived~\cite{KGE13}. 
Finally, there are also tree topologies at N$^4$LO (Fig.~\ref{fig_3nfn4lotrees}) which include
a new set of 3N contact interactions (graph (c)). These 3N contacts have recently been derived
by the Pisa group~\cite{GKV11}. Contact terms are typically simple (as compared
to loop diagrams) and their coefficients are unconstrained (except for naturalness). 
Therefore, it would be an
attractive project to test some terms (in particular, the spin-orbit terms) 
of the N$^4$LO contact 3NF~\cite{GKV11} in calculations of few-body reactions (specifically,
the p-d and p-$^3$He $A_y$) and spectra of light nuclei.

\section{Conclusions and Outlook}

The past 15 years have seen great progress in our understanding of nuclear forces
in terms of low-energy QCD. Key to this development was the realization that
low-energy QCD is equivalent to an effective field theory which allows for 
a perturbative expansion that has become known as chiral perturbation theory.
In this framework, two- and many-body forces emerge on an equal footing and the empirical fact
that nuclear many-body forces are substantially weaker then the two-nucleon force
is explained automatically.

In spite of the great progress and success of the past 15 years, there are still some
unresolved issues. One problem is the
proper renormalization of the chiral two- and many-nucleon potentials, where systematic
investigations are already under way (cf.\ Sec.~\ref{sec_reno}).

The other unfinished business are the few-nucleon forces beyond NNLO (``sub-leading
few-nucleon forces'') which are needed to hopefully resolve some important outstanding
nuclear structure problems. At orders N$^3$LO and N$^4$LO very many
new 3NF structures appear, some of which have already been tested.
However, in view of the multitude of 3NF topologies it will take a while until
we will have a proper overview of impact and convergence of these contributions.

If the open issues discussed in this paper will be resolved within
the next few years, then, after 70 years of desperate struggle, we
may finally claim that the nuclear force problem is essentially under control.
The greatest beneficiaries of such progress will be the 
{\it ab initio} nuclear structure physicists, including {\it James Vary}.
May this be a birthday present for him. 
\\
\\
This work was supported in part by the U.S. Department of Energy
under Grant No.~DE-FG02-03ER41270.

\end{document}